# COAXIAL WIRE MEASUREMENTS IN NLC ACCELERATING STRUCTURES†

N. Baboi*, R.M. Jones, G.B. Bowden, V. Dolgashev, S.G. Tantawi and J.W. Wang

†Stanford Linear Accelerator Center,
2575 Sand Hill Road, Menlo Park, CA, 94025
*On leave from NILPRP, P.O. Box MG-36, 76900 Bucharest, Romania

## Abstract

The coaxial wire method provides an experimental way of measuring wake fields without the need for a particle beam. A special setup has been designed and is in the process of being fabricated at SLAC to measure the loss factors and synchronous frequencies of dipole modes in both traveling and standing wave structures for the Next Linear Collider (NLC). The method is described and predictions based on electromagnetic field simulations are discussed.

*Paper presented at the 2002 8th European Particle Accelerator Conference (EPAC 2002)
Paris, France,
June 3rd -June 7th, 2002*

This work is supported by Department of Energy grant number DE-AC03-76SF00515†

# COAXIAL WIRE MEASUREMENTS IN NLC ACCELERATING STRUCTURES†

N. Baboi[*], R.M. Jones, G.B. Bowden, V. Dolgashev, S.G. Tantawi and J.W. Wang,
SLAC, Stanford, CA 94309

*Abstract*

The coaxial wire method provides an experimental way of measuring wake fields without the need for a particle beam. A special setup has been designed and is in the process of being fabricated at SLAC to measure the loss factors and synchronous frequencies of dipole modes in both traveling and standing wave structures for the Next Linear Collider (NLC). The method is described and predictions based on electromagnetic field simulations are discussed.

## 1 INTRODUCTION

If left undisturbed, the wake fields excited by bunches of highly energetic charged particle beams interact with trailing bunches and can lead to a significant dilution of the beam emittance. Provided the wakefield is sufficiently strong, it can cause a Beam Break Up (BBU) instability making the collider unusable. The final center of mass energy of the beam for the NLC is 500 GeV (in the first stage of operation) and should the beam become unstable and come into contact with the metallic structure it would readily vaporize the metal. This makes it crucially important to understand and minimize the wakefield in the design of linear colliders.

Transverse wake fields, for both the first dipole band and higher order bands, have been carefully measured at the ASSET facility in the SLC [1] for several DDS accelerators. In that experiment, the wakefields excited by a bunch were sampled with a subsequent witness bunch. Also, the fields radiating out from the manifolds were sampled and information on the modal content and beam position was obtained via coaxial pickups.

In this paper a wire method is described. The advantage of this method lies in the fact that it does not require the use of particle beams [2] to measure the wakefield. A microwave measurement in the laboratory allows the wakefield to be measured in a cost-effective manner. In essence, the method relies on being able to simulate a beam passing through the structure by using a microwave signal traveling on a wire placed at varying offsets from the electrical center of the structure. This microwave signal excites a whole range of higher order modes. One can measure either the distortion of a current pulse at the exit of the structure in time domain [2], or the transmission parameter $S_{21}$ as a function of frequency from which the impedance of the higher order modes and passbands can be deduced [3,4]. Integrating the impedance one obtains the loss factors.

Such measurements have been routinely used for determining the impedance of accelerating structures and other components [5,6,7]. Either an on-axis wire for longitudinal impedances, or two wires placed symmetrically around the axis for transverse impedances, have been used [8]. The measurement is more difficult to perform for the NLC X-band structures, compared to some earlier lower frequency measurements, because of rather tight tolerances imposed on component fabrication by an accelerator operating at a fundamental mode of frequency of 11.424GHz. Special attention has also to been paid to the alignment of components in the measurement setup.

## 2 EXPERIMENTAL SETUP

For the NLC structures, frequency domain measurements will be performed first. We are primarily interested in the kick factors of the dipole passbands that give the main contribution to the wake fields [9]. In order to excite and measure such fields, we need either an offset wire, or two wires. Initially, we will use a single wire. A discussion of these two options is given in [10].

The measurement setup is shown in Fig. 1. With the aid of a network analyzer the transmission parameter $S_{21}$ is measured through the structure, which will be referred to as a device under test (DUT). The matching sections were carefully designed such that signal losses are minimized. In order to connect the wire relatively easily, we chose to use coaxial towaveguide adapters, as shown in Fig. 2. Such adapters are characterized by a broad frequency bandwidth, enabling us to measure one or possibly two pass-bands in each experiment. The first adapter (left in the figure) is commercially available, while the second one was designed to accommodate the wire and to adapt the inner conductor of the coaxial waveguide to the wire diameter. The wire diameter was chosen 0.25 mm. For thinner wires the losses of the coaxial waveguide increase, while for thicker wires the perturbation of the fields in the structure is higher. The adapter used for the lowest dipole bands of the accelerating structures has a frequency range from 11 to 18 GHz. It suppresses the reflected RF signal below approximately 30 dB. The wire can be transversely moved with respect to the axis by slightly bending the tubes of the matching sections. This solution was chosen to avoid RF discontinuities in the setup.

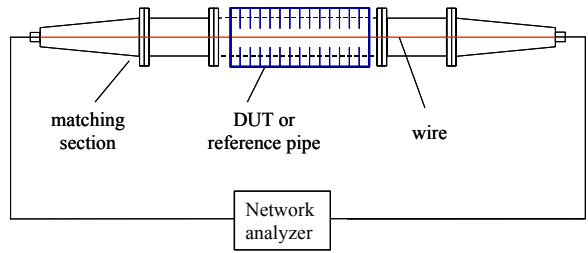

Figure 1 Setup of the wire measurement

† Supported by the DOE, grant number DE-AC03-76SF00515
* On leave from NILPRP, P.O. Box MG-36, 76900 Bucharest, Romania

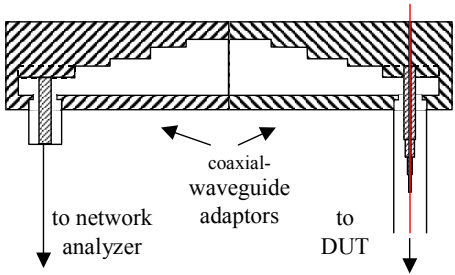

Figure 2 Matching section

Two measurements are required to properly characterize the DUT, one with the DUT present and one with a reference pipe only. This allows the longitudinal loss factor to be calculated and this enables the kick factor to be obtained.

## 3 SIMULATIONS

In order to understand the changes induced by the presence of the wire, an analysis of an accelerating structure has been made. We used the finite element electromagnetic field simulator HFSS to simulate the fields in the structure and to calculate the resonant modes. The detuned accelerating structure DS2 [11] was used for initial studies because it has an adequate frequency separation of modes and is readily available for testing in the wire measurement set up. Simulations have been made for a downstream cell of this structure, with periodic boundary conditions and work is under way to simulate the progress of a pulse through the DUT.

### 3.1 Dispersion diagram

Fig. 3 shows the dispersion diagram obtained for cell 198 of DS2. The dashed lines with empty symbols represent passbands and modes for the unperturbed cell. A quarter of cell was simulated (see Fig. 4-a) for two cases: with magnetic boundary conditions in both symmetry planes and with electric boundary in one plane and magnetic in the other. The walls are perfect conducting. The solid lines with solid symbols in Fig. 3 are obtained for a perfect conducting wire 250 μm in diameter placed at 1 mm distance from the axis and parallel to it. Half of a cell was simulated (Fig. 4-b) with magnetic boundary conditions. The first three dipole bands are included, as well as their equivalent for the perturbed case [12]. The synchronous dipole modes are also marked on the figure for both the unperturbed and the perturbed cases.

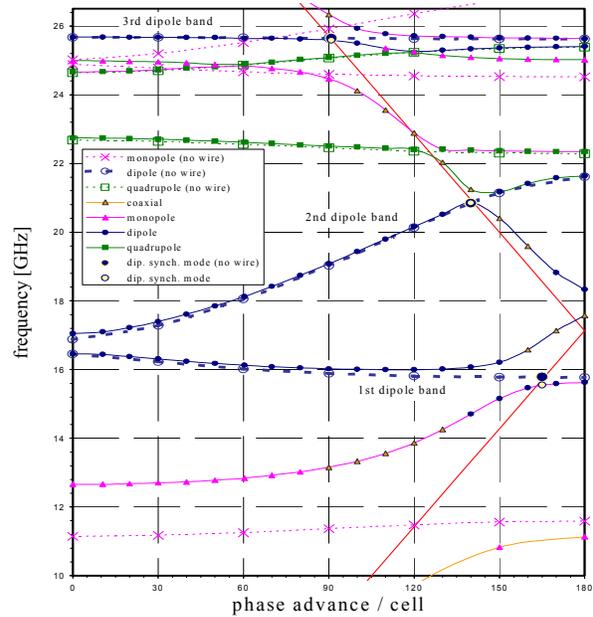

Figure 3 Dispersion diagram for cell 198 of DS2, with wire displacement equal to 1 mm (solid lines and full symbols) compared to the unperturbed case (dashed lines and empty symbols)

The avoided crossings indicate couplings between the passbands, e.g. between the coaxial TEM mode, following the light velocity line, and the first monopole band around 12 GHz and 130º phase advance, and between the first dipole band and the monopole band around 16 GHz and 160º. An immediately observed effect on the synchronous dipole modes is the shift in frequency. It converges to the frequency of the unperturbed mode when the wire approaches the center of the cell.

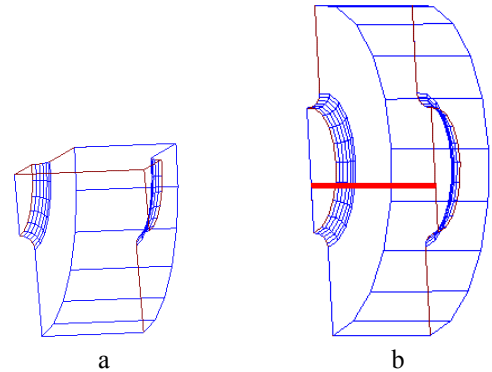

Figure 4 Cell 198 of the DS2 structure, as used for simulation of unperturbed cell (a), and for a single wire placed off-axis (b)

## 3.2 Kick factors

The loss factor for the cell with wire is obtained as usual from integrating the longitudinal electric field along a path parallel to the axis. For dipole modes, the field has a term proportional to the radial coordinate, *r*, (measured with respect to the wire position) as in the case when no wire is present in the cell, plus a term proportional to $1/r$. The second term dominates for small offsets, and it becomes negligible for larger offsets. This entitles us to define a kick factor as in the case of a cell without wire, by normalizing the loss factor to the offset of the path where the electric field is integrated:

$$k_\perp = \frac{c}{\omega} \cdot \frac{1}{(x-x_w)^2 \cdot l_{cell}} \cdot k_\parallel$$

where *c* is the velocity of light, ω the angular frequency of the mode, $l_{cell}$ the cell length, x is the offset of the integration path considered in the plane containing the wire, $x_w$ the offset of the wire and $k_\parallel$ the loss factor calculated along the path at offset x.

Fig. 5 shows the kick factor for the first dipole mode for a wire placed at 1 mm offset from the axis as a function of the offset of the integration path with respect to the wire. The kick factor for the case without a wire is shown as well. For large offsets with respect to the wire, the kick factor is constant. The asymmetry between positive and negative offsets is a result of the asymmetry in the fields.

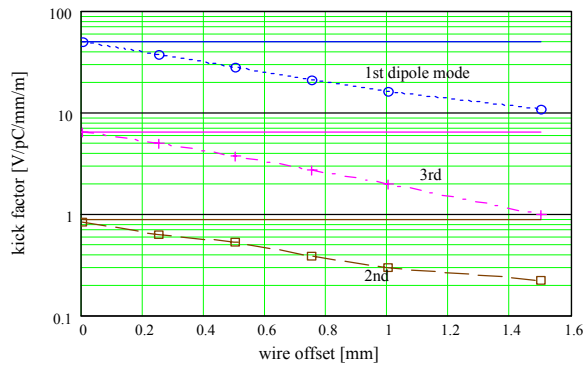

Figure 6 Kick factors of the first 3 dipole synchronous modes of cell 198 of DS2 with a wire 250 μm in diameter placed parallel to the axis, as a function of the wire offset

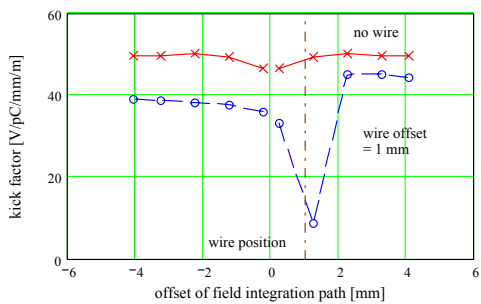

Figure 5 Kick factor for the first synchronous dipole mode of cell 198 with a wire 250 μm in diameter, placed at 1 mm offset

The change in the dipole mode kick factor with the wire offset is illustrated in Fig. 6. As the wire approaches the center of the cell, the kick factor becomes equal to the one of the unperturbed cell (marked by the horizontal lines). In practice one cannot excite the dipole modes with a centered wire, but one can deduce the synchronous frequency and kick factor of the mode by measuring with several wire offsets and extrapolating to centered wire.

## 4 SUMMARY

In this paper we have described the experimental setup and the fundamental physics that lie behind the wire measurements which will be used to measure wakefields in the NLC accelerating structures. The experiment has been designed to minimize transmission losses in the matching sections. Furthermore, the transverse displacement system envisaged is straightforward, and does not cause unnecessary spurious mode excitation.

We have demonstrated that in the limit of the wire offset going to zero the beam kick factors and synchronous frequencies are obtained.

The various components of the setup are in the process of being put together and testing of the setup is expected in the near future.

## 5 ACKNOWLEDGEMENTS

We thank the structures group at SLAC, where these results were first presented, for providing stimulating discussions and we would like to the thank P.B. Wilson, R.H. Miller and R.D. Ruth for particularly useful discussions.